\documentclass[12pt,aps,prd,preprint,tightenlines,superscriptaddress,
  showpacs,nofootinbib]{revtex4}
\newcommand{\PRE}[1]{{#1}} 

\usepackage{hyperref}      
\usepackage{bm}
\usepackage{epsfig}
\usepackage{amsmath}
\usepackage{array}
\renewcommand{\eqref}[1]{Eq.~(\ref{#1})}

\newcommand{\secref}[1]{Sec.~\ref{sec:#1}}

\newcommand{\bea}{\begin{eqnarray}}
\newcommand{\eea}{\end{eqnarray}}
\newcommand{\beq}{\begin{equation}}
\newcommand{\eeq}{\end{equation}}
\newcommand{\beqa}{\begin{eqnarray}}
\newcommand{\eeqa}{\end{eqnarray}}
\newcommand{\nn}{\nonumber}

\newcommand{\drawsquare}[2]{\hbox{%
\rule{#2pt}{#1pt}\hskip-#2pt
\rule{#1pt}{#2pt}\hskip-#1pt
\rule[#1pt]{#1pt}{#2pt}}\rule[#1pt]{#2pt}{#2pt}\hskip-#2pt
\rule{#2pt}{#1pt}}

\newcommand{\Yfund}{\drawsquare{6.5}{0.4}}
\newcommand{\Yafund}{\overline{\Yfund}}
\newcommand{\Yasymm}{\raisebox{-3.5pt}{\drawsquare{6.5}{0.4}}\hskip-6.9pt%
         \raisebox{3pt}{\drawsquare{6.5}{0.4}}}
\newcommand{\Ysymm}{\drawsquare{7}{0.6}\hskip-0.6pt%
	\drawsquare{7}{0.6}}

\newcommand{\Yadjoint}{\raisebox{-3.5pt}{\drawsquare{6.5}{0.4}}\hskip-6.9pt%
        \raisebox{3pt}{\drawsquare{6.5}{0.4}}\hskip-0.4pt
        \raisebox{3pt}{\drawsquare{6.5}{0.4}}}

\newcommand{\barq}{\bar{Q}}

\newcommand{\muv}{M_{{\rm UV}}}
\newcommand{\ruv}{r_{{\rm UV}}}

\newcommand{\oraf}{O'Raifeartaigh}

\newcommand{\headmath}[1]{\texorpdfstring{$#1$}{#1}}

\newcommand{\vev}[1]{\langle #1\rangle}
\begin{document}

\title{ \PRE{\vspace*{1.5in}} Singlet-Assisted Supersymmetry Breaking
\\
\PRE{\vspace*{0.3in}} }

\author{Yael Shadmi\PRE{\vspace*{.5in}}}
\affiliation{Physics Department, Technion---Israel Institute of
Technology,\\ Haifa 32000, Israel
\PRE{\vspace*{.5in}}
}
\author{Yuri Shirman\PRE{\vspace*{.5in}}}
\affiliation{Department of Physics and Astronomy, University of
California, Irvine, CA 92697, USA
\PRE{\vspace*{.2in}}
}

\date{October 2011}

\begin{abstract}
\PRE{\vspace*{.3in}} 
We describe a simple recipe for obtaining local supersymmetry-breaking
vacua in s-confining theories coupled to gauge singlets.
This recipe gives rise to effective O'Raifeartaigh models
in the IR, with calculable supersymmetry-breaking
minima near the origin, and can be applied to both
vector-like and chiral theories.
Since the properties of the vacuum are largely determined by
superpotential terms that are non-renormalizable in the UV,
it is calculable even when all dimensionless
couplings are taken to be of order one.
By construction, the models preserve a large subgroup of the original
global symmetry. 
While we only study here s-confining theories,
we expect our tools  to be useful for inducing dynamical supersymmetry 
breaking in many gauge theories. 
\end{abstract}


\maketitle

\section{Introduction}
\label{sec:introduction}
Dynamical supersymmetry breaking (DSB) provides a beautiful
mechanism for generating the hierarchy between the weak scale
and the Planck scale~\cite{Witten:1981kv}. 
One of the least satisfying aspects of this idea is the lack of a general 
organizing principle for finding DSB theories~\cite{reviews}. 
This is related to the fact that these theories are 
non-generic.
It is reasonable to expect this problem to be ameliorated if
one does not insist on global supersymmetry-breaking minima,
and instead only searches for local minima.
Indeed, local DSB minima were found to occur in some of the simplest 
supersymmetric gauge theories, namely supersymmetric QCD~\cite{iss},
suggesting that local DSB vacua are quite generic.
Furthermore, if one only requires a local minimum,
it may not be necessary to lift all of the classical flat directions,
so that the classical superpotential one adds can be less intricate,
and the resulting models more generic.

Simpler superporpotentials have another potential advantage. 
The classical superpotential necessarily breaks some of the
global symmetry of the original theory.
This is problematic for constructing models of direct gauge mediation,
since these are obtained by embedding the standard model gauge group
in a (weakly gauged) subgroup of the global symmetry of the DSB 
model~\cite{Giudice:1998bp}.
If one is not forced to lift all the classical flat directions,
a larger subgroup of the original global symmetry is likely
to survive.

In this paper, we describe a simple recipe for obtaining 
DSB minima in many s-confining theories~\cite{Seiberg:1994bz,Intriligator:1995ne,pouliot,sconf}, 
by coupling these theories to gauge singlets.
Essentially, the construction yields an effective O'Raifeartaigh
model in the IR, with a supersymmetry breaking vacuum near the
origin of the moduli space, much like in the models of~\cite{iss}.
Our recipe can be applied to both vector-like theories,
such as s-confining SQCD, 
and to chiral theories (for metastable susy breaking in chiral theories without singlets see \cite{Lin:2011vd,Shadmi:2011mt}).
It is also closely related to the models of~\cite{it,iy}
in which DSB is achieved through couplings to gauge singlet
fields. 

In order to obtain the supersymmetry-breaking minimum
near the origin, we do not need to lift all of the classical
flat directions. Just as in~\cite{it,iy}, the classical
moduli space is parametrized by the gauge singlets.
The fate of this classical moduli space is somewhat model
dependent. As we will see, in some of the s-confining theories 
it is lifted non-perturbatively like in the models of ~\cite{it,iy},
while in others, for large field VEVs, there are runaway directions
along which the potential asymptotes to zero.
In any case, the supersymmetry-breaking vacuum near the origin is
separated from these by a non-calculable potential barrier.

Specifically, we will introduce singlet fields $S_I$ for each of the gauge
invariants $O_I$ of the s-confining theory, apart from a single one.
Denoting this last gauge invariant by $O_0$, we add
the superpotential\footnote{The addition of a tadpole for a gauge 
invariant modulus is required to exclude the supersymmetric minimum 
at the origin of the field space. 
In the models of \cite{it,iy} such a tadpole was not necessary since 
the origin did not belong to the quantum moduli  space.},
\beq\label{basicw}
W= S_I O_I + O_0\ .
\eeq
Below the confinement scale of the theory, the last term of~\eqref{basicw}
becomes a tadpole. 
Since the non-perturbative dynamics of s-confining theories 
often gives rise to superpotential terms that are cubic in the gauge 
invariants, 
the IR theory reduces to an O'Raifeartaigh model. 
In some cases, such as s-confining SQCD, all the IR fields
appear in the O'Raifeartaigh superpotential.
In more complicated examples, the gauge invariants separate into two sets.
One set constitutes an O'Raifeartaigh model, while the second
set consists of massive fields whose only superpotential interactions
are non-renormalizable (NR).
Since these latter fields obtain masses in the IR through the 
superpotential~\eqref{basicw}, their vacuum expectation values (VEVs) 
vanish and they do not affect the supersymmetry-breaking minimum.

Most of the gauge invariants $O_I$ have dimensions larger than 2.
The coefficients of the corresponding terms in~\eqref{basicw} are then
naturally small.
In particular, choosing $O_0$ to be the highest dimension gauge invariant
results in a calculable minimum, without having to take any order-one
parameters to be small.

We only study s-confining theories in this paper.
These are particularly simple to analyze since
their  moduli spaces are parametrized by the 
gauge invariants $O_I$~\cite{Seiberg:1994bz}. 
Furthermore, all of these theories have been found
and their superpotentials are known
in~\cite{Seiberg:1994bz,sconf,Spiridonov:2009za}.
However, we expect this construction
to be applicable in a wide variety of models, including models with 
IR-free dual descriptions.

The construction described above gives rise to the simplest
O'Raifeartaigh models in the IR, and therefore leaves the
$R$-symmetry unbroken. 
As we will show, however, slight variations of these models
give O'Raifeartaigh-type models with some fields
having $R$-charges different from 0 or 2
as in~\cite{Shih:2007av}, leading to $R$-breaking minima. 
This requires taking 
some couplings that would naturally be order one, to be small.

As mentioned above, one of the inherent difficulties 
in using DSB models in the context of direct gauge mediation 
is that the superpotential required for lifting the flat directions
breaks some of the global symmetry. 
The recipe we described above is clearly advantageous
from this point of view, since the $S_I O_I$ terms do not break
any symmetries. In fact, the $S_I$'s and $O_I$'s are in conjugate 
representations of the global symmetry, and can therefore be 
used as vector-like messenger
pairs once the standard-model subgroup of that global symmetry is gauged.

This paper is organized as follows.
In~\secref{even}, we study in detail the even-$n$~SU($n$) theories
with a single anti-symmetric tensor and four flavors.
We will show that these have a calculable supersymmetry-breaking minimum
near the origin, and no supersymmetric minima for VEVs smaller
than the cutoff scale.
We will also present an $R$-symmetry breaking modification
of the basic model.
We generalize this construction to the even-$n$ theories in~\secref{even}.
The remaining s-confining SU($n$) theories are discussed in~\secref{general}.

\section{The even-\headmath{n} single anti-symmetric tensor SU(\headmath{n}) theories}
\label{sec:even}
As a first example, we take SU($2N$), with a single antisymmetric
tensor, and four flavors. This is a chiral theory---only 
four pairs of fundamentals and anti-fundamentals can be given mass. 
The field content of the theory is summarized in the first part
of Table~\ref{mattercontent}.
\begin{table}
\begin{tabular}{|>{$\displaystyle}c<{$}|>{$\displaystyle}c<{$}|>{$\displaystyle}c<{$}|>{$\displaystyle}c<{$}|>{$\displaystyle}c<{$}|>{$\displaystyle}c<{$}|>{$\displaystyle}c<{$}|>{$\displaystyle}c<{$}|} \hline
 &SU(2N)&SU(2N)&SU(4)&U(1)^\prime&U(1)&U(1)_R&{\rm dim}\\
\hline
\vphantom{\Yasymm^2}
A & \Yasymm &1&1&0&4&0&1\\
\bar Q& \Yafund &\Yfund & 1& 4& 0&1/N&1\\
Q&\Yfund &1&\Yfund&-2N&-2(N-1)&0&1\\
\hline
\hline
\vphantom{\bar Q^{Q^{Q^a}}}
M_{ia}\sim(Q_i\bar Q_a)&&\Yfund&\Yfund&4-2N&-2(N-1)&1/N&2\\
\vphantom{\bar{Q}^{2^2}}
P_{ab}\sim(A\bar{Q}^2)&&\Yasymm&1&8&2&2/N&3\\
\hline
\vphantom{A^{(N)^{2^2}}}
Y_{ij}\sim(A^{(N-1)}Q^2)&&1&\Yasymm&-4N&0&0&N+1\\
\vphantom{A^{N^{2^2}}}
X\sim(A^{N-2}Q^4)&&1&1&-8N&-4N&0&N+2\\
\vphantom{A^{N^{2^2}}}
B\sim(\bar Q)^{2N}&&1&1&8N&0&2&2N\\
\vphantom{A^{N^{2^2}}}
\tilde A\sim (A^N)&&1&1&0&4N&0&N\\
\hline
\hline
\vphantom{A^{N^{2^2}}}
S_M^{ia}&&\bar{\Yfund}&\bar{\Yfund}&-4+2N&2(N-1)&2-1/N&1\\
\vphantom{A^{N^{2^2}}}
S_P^{ab}&&\bar{\Yasymm}&1&-8&-2&2-2/N&1\\
\vphantom{A^{N^{2^2}}}
S_Y^{ij}&&1&\bar{\Yasymm}&4N&0&2&1\\
\vphantom{A^{N^{2^2}}}
S_X&&1&1&8N&4N&2&1\\
\vphantom{A^{N^{2^2}}}
S_A&&1&1&0&-4N&2&1\\
\hline
\end{tabular}
\caption{Matter content of the $SU(2N)$ model. The quantum numbers for 
charged elementary fields are given in the top part, 
for gauge invariant composites in the middle part, 
and for gauge singlets in the bottom part.}
 \label{mattercontent}
\end{table}
This theory was studied in~\cite{pouliot,sconf}.
We have listed in the Table \ref{mattercontent} the anomaly-free global symmetry
of the theory before the addition of any tree-level superpotential.

While this might seem as a complicated first example to choose,
we will see that the fields of the IR theory ``factorize'' into 
two sets. 
One set only appears in those terms of the dynamical superpotential 
that remain non-renormalizable in the IR  
and is therefore irrelevant 
for studying the potential near the origin, 
while the second set constitutes
an O'Raifeartaigh model.
This factorization is generic to the s-confining theories.

The IR theory can be described in terms
of the gauge invariants defined in the second part of 
Table~\ref{mattercontent},
\beq\label{gaugeinv}
M_{ia},\ \  P_{ab},\ \  Y_{ij},\ \  X,\ \ 
\tilde{A}, \ \  B, 
\eeq
where $a,b=1,\ldots,2N$ and  $i,j=1,\ldots,4$,
with the dynamically generated superpotential
\beq\label{wdyn}
 W_{dyn}=
\frac{1}{\Lambda^{4N-1}}
\,\left\{\left[\tilde{A}M^4 P^{N-2}+ Y M^2 P^{N-1}
+X P^N\right] + \left[B \left(\tilde{A} X+Y^2\right)\right]\right\}\ .
\eeq
Note that $M$ and $P$ only appear in the first three terms
of this superpotential, which are non-renomalizable
and involve high powers of these fields.
The last two terms involve $A$, $X$, $Y$, and $B$ only,
and will become a part of an O'Raifeartaigh model leading to 
supersymmetry breaking.
 Here and in the following, we separate the superpotential
into two parts, to highlight this factorization

\subsection{Classical superpotential and flat directions}\label{classtheory}
In order to obtain the basic O'Raifeartaigh model 
we add singlets for each one of the gauge invariants
apart from $B$ (see bottom part of Table~\ref{mattercontent}),
with the superpotential,
\begin{equation}\label{w00}
\begin{split}
W_0 = &\left[\lambda_M 
 S_M^{ia} (Q_i\bar Q_a) 
+ \frac{\lambda_P}{\muv}  S_P^{ab} (A\bar{Q}^2)_{ab} \right]+\\
&\left[
\frac{\lambda_A}{\muv^{N-2}}  S_A   (A^N) +
\frac{\lambda_{Y}}{\muv^{N-1}}   S^{ij}_Y  (A^{(N-1)}Q^2)_{ij}
+\frac{\lambda_X}{\muv^N}   S_X  (A^{N-2}Q^4)
 +\frac{\lambda_B}{\muv^{2N-3}} (\bar Q)^{2N}
\right] 
\end{split}
\end{equation}
where  parantheses denote the contraction of SU($2N$) indices.
The $\lambda$'s are order-one couplings. We could in principle
absorb these into the singlets but it would be convenient to
keep them explicit for the discussion below.
This superpotential lifts all the flat directions associated with the 
charged fields. The classical moduli space is therefore
parameterized by the singlets. In Section~\ref{susybreaking} we will 
argue that near the origin of the moduli space all flat directions are 
lifted and a supersymmetry-breaking local minimum exists.
In Section \ref{susyminima} we will show that at large VEVs most of 
the singlet classical flat directions are lifted at the quantum level 
while the only remaining runaway along the $S_P$ direction is 
separated by a potential barrier from the supersymmetry breaking minimum.

The superpotential~\eqref{w00}
is not the most generic one consistent with the 
unbroken SU(4)$\times$SU(2$N$)$\times$U(1)$\times$U(1)$_R$~global symmetry
(U$(1)^\prime$ is broken explicitly).
However, the only additional tree level terms that can be added to~\eqref{w00}
that are consistent with this symmetry
are suppressed compared to terms included in \eqref{w00} both in the UV and in
the IR.
Indeed, new operators 
are obtained by multiplying the existing terms by $AX$ or by the 
classically equivalent combination $Y^2$.
In the UV these operators start at dimension  $2N+5$, whereas the highest 
dimension term of~\eqref{w00} has dimension $2N$. 
In the IR, these operators start at dimension 4,
while the maximal IR dimension of the terms in~\eqref{w00} is 2.

\subsection{The supersymmetry-breaking vacuum}
\label{susybreaking}

We will now show that there is a local supersymmetry-breaking 
minimum near the origin\footnote{This analysis is very similar
to the analysis of~\cite{Chacko:1998si}.}.
For finite field VEVs, the K\"ahler potential in terms of the gauge invariants
is non-singular, so that these gauge invariants can be
used to study the IR theory.
The last term of~\eqref{w00} is then linear in the baryon $B$,
and the remaining terms become mass terms near the origin.
The full superpotential is given by,
\beqa\label{exactw}
 W&=&\tilde{A}M^4 P^{N-2}+ Y M^2 P^{N-1}
+X P^N + B \left(\tilde{A} X+Y^2\right)\nn\\
&+& m_M  S_M^{ia} M_{ia} + m_P S_P^{ab} P_{ab}
+m_Y S^{ij}Y_{ij}
+ m_X S_X X +m_A S_A \tilde{A}+ f_B B\ ,
\eeqa
with the masses
\beq\label{masses}
m_M= \lambda_M \,\Lambda,~~m_P=\lambda_P\, \ruv \,\Lambda,~~
m_A=\lambda_A \,\ruv^{N-2}\, \Lambda,~~
m_{Y}=\lambda_Y\, \ruv^{N-1}\, \Lambda,~~
m_X=\lambda_X \,\ruv^{N} \,\Lambda,
\eeq
and
\beq\label{fb}
f_B=\lambda_B \,\ruv^{2N-3}\, \Lambda^2,
\eeq
where we defined
\beq
\ruv \equiv\frac\Lambda\muv \ll 1 . 
\eeq
The masses~\eqref{masses} are hierarchical, with
\beq
m_M\sim\Lambda \gg m_P \gg m_A \gg  \sqrt{f_B}\gg m_{Y}\gg m_X,
\eeq
and
\beq\label{ratios}
m_x m_A \sim m_Y^2 < f_B \ .
\eeq
It is easy to see that there are no supersymmetric minima at finite VEVs.
The singlet $F$-term equations set all invariants to zero except $B$.
The $B$ equation is schematically
\beq
AX + Y^2 + f_B=0
\eeq
and this cannot be satisfied for $A=X=Y=0$.

Near the origin, 
we can integrate out the heaviest fields, $M$, $P$ and the
corresponding singlets, 
to get an effective O'Raifeartaigh
model with the superpotential,
\beq
 W= B \left(\tilde{A} X+Y^2 + f_B  \right) +
m_Y S^{ij}Y_{ij} + m_X S_X X +m_A S_A A \ .
\eeq
The behavior of this model is well known.
Supersymmetry is broken since the $S_A$, $S_X$,
and $S_Y^{ij}$ equations set $A$, $X$ and $Y_{ij}$
to zero respecively, in conflict with the $B$ equation.
At tree-level, there is a flat non-zero potential
with $B$ undetermined. This potential is corrected
radiatively, and the one-loop Coleman-Weinberg potential
leads to a minimum at $B=0$, 
with small or zero VEVs for the remaining fields. 
To see this, recall that with no tuning
of $\lambda_{A,X,Y}$, we have 
$m_A m_X\sim m_Y^2$.
To simplify the problem, we can study the model in two
limits. Take first $\lambda_A< \ruv$, so that  $m_A<m_Y$.
The $Y$ spectrum is then approximately supersymmetric, and we can integrate
out $Y$ and $S_Y$, recovering the simplest O'Raifeartaigh
model, with a minimum at 
\beqa\label{minax}
B&=&0\,,\ \ \ S_X=S_A=0\,,\\
A&=&\sqrt{\frac{m_X}{m_A} (f_B-m_A m_X)}\,,~~
X=\sqrt{\frac{m_A}{m_X} (f_B-m_A m_X)}\,.
\eeqa
Taking instead $\lambda_Y<\ruv$ so that $Y$ is lighter than
$A$, $X$, the minimum is at 
\beq\label{miny}
B=0\,,~~S_Y=0\,,~~ 
Y=\sqrt{f_B-m_Y^2}\,.
\eeq
In both of these limits,
the $B$ mass is given by
\beq\label{mbcw}
{m_B^2}^{(CW)}  = \frac1{16\pi^2} \, \frac{m_1^2m_2^2}{f_B}\ 
\eeq
with $m_1 m_2=m_X m_A$ in the former case ($\lambda_A< \ruv$), 
and $m_1 m_2=m_Y^2$
in the latter ($\lambda_Y<\ruv$).
Restoring $\lambda_A\sim\lambda_Y\sim1$, it is clear then
that~\eqref{mbcw} still holds, while the VEVs of $A$, $X$ and $Y$
are somewhere between the values of~\eqref{minax},~\eqref{miny} and zero.

All these VEVs are parametrically smaller than $\Lambda$.
Using~\eqref{masses} we can find for the VEVs of $A$ and $X$ of~\eqref{minax} 
\beqa\label{minax1}
A \sim \lambda_A^{-1/2} \,\ruv^{N-1/2}\,  \Lambda \ll \Lambda\nn\\
X\sim \lambda_A^{1/2} \,\ruv^{N-5/2}\,  \Lambda \ll \Lambda\nn
\eeqa
and for the $Y$ VEV of~\eqref{miny}
\beq\label{miny1}
Y \sim \ruv^{N-3/2}\,  \Lambda \ll \Lambda\ .
\eeq

\subsection{Calculability}
We can now see why the analysis above is reliable.
Near the origin, 
the K\"ahler potential is roughly of the form
\beqa
K&=& 
M^\dagger M \left[1+g_M\left(\frac{M}\Lambda,..,
\frac{\lambda_M S_M}\Lambda\right)\right] 
+ B^\dagger B  \left[1+g_B\left(\frac{M}\Lambda,..,
\frac{\lambda_M S_M}\Lambda\right)\right]\nn\\
&+& S^\dagger S \left[1+\lambda_M g_{S_M}\left(\frac{M}\Lambda,..,
\frac{\lambda_M S_M}\Lambda\right)\right] +\ldots
\eeqa
where $g_M$ etc are non-calculable functions,
and the ellipses stand for terms involving the remaining fields.

Since there are mass terms for $M$, $P$, $S_M$, $S_P$ near the origin,
these fields have zero VEVs and zero $F$-terms at the
minimum of interest, so the non-calculable K\"ahler terms 
involving these fields can be neglected.
Likewise, the NR part of the dynamical superpotential involves
at least three powers of $M$ and $P$, and therefore does not
generate a potential near the origin.

As we saw above, the fields $A$, $X$, $Y$ all have vanishing $F$-terms
and VEVs much smaller than $\Lambda$, so higher-dimension K\"ahler terms
involving these fields can be neglected.
The only potentially dangerous K\"ahler terms involve $B$, 
which has a non-zero $F$-term, and therefore generates a potential
\beq
V = f_B^2(1+ \alpha B^\dagger B/\Lambda^2 +\ldots)
\eeq
where $\alpha$ is a non-calculable order-one coefficient, and the ellipsis
stands for similar terms involving the other fields.
Thus, there are non-calculable contributions to the masses-squared
that scale as\footnote{The corresponding non-calculable contribution
to the fundamental singlet $S_I$ additionally involves at least two powers
of the corresponding singlet coupling $\lambda_I$.}
\beq\label{noncalc}
m^2_{non-calc}\sim f_B^2/\Lambda^2 \sim \ruv^{4N-6}\, \Lambda^2\,.
\eeq
This contribution is small compared to the tree-level
masses of all the fields in the model.
The pseudo-modulus $B$, whose calculable mass only arises from 
the Coleman-Weinberg 
potential, requires a more detailed estimate. 
The radiatively generated $B$ mass depends on the relative sizes of
$f_B$ and $m_1 m_2$, where $m_1$, $m_2$ are the masses defined in~\eqref{mbcw}.
If $m_1 m_2 > f_B$ we have
\beq\label{smallfb}
{m_B^2}^{(CW)}  \sim \frac1{16\pi^2} \, f_B 
\eeq
so that
\beq\label{noncalcratio}
\frac{m^2_{non-calc}}{{m_B^2}^{(CW)}} \sim 
{16\pi^2}\, \frac{f_B}{\Lambda^2} \,
\eeq
which is clearly small.

If however, as is the case in our model,
$m_1 m_2 < f_B$, 
we have
\beq
{m_B^2}^{(CW)}  = \frac1{16\pi^2} \, \frac{m_1^2m_2^2}{f_B}\,, 
\eeq
so that
\beq\label{noncalcratio1}
\frac{m^2_{non-calc}}{{m_B^2}^{(CW)}} \sim 
{16\pi^2}\, \frac{f_B^3}{\Lambda^2 m_1^2 m_2^2} \,,
\eeq
and whether or not this is smaller than one depends on the ratio 
$f_B/(m_1m_2)$.
Using the parametric dependences of the various scales on $\ruv$
of~\eqref{masses} we find 
\beq\label{noncalcratio2}
\frac{m^2_{non-calc}}{{m_B^2}^{(CW)}} \sim 
{16\pi^2}\, \ruv^{2N-5}\,,
\eeq
so that for small $\ruv$ the non-calculable contribution is subdominant 
in this case as well.

We can also see now why we chose the highest-dimension gauge invariant
for the supersymmetry breaking tadpole. 
The O'Raifeartaigh model with
\beq
W= \phi ( \phi_1 \phi_2 + f) + m_1 S_1\phi_1 + m_2  S_2\phi_2e
\eeq
has a minimum with $\phi_{1,2}=0$ for $f< m_1 m_2$.
Thus if we can achieve  $f< m_1 m_2$, the minimum is at the origin
and the  pseudo-modulus $\phi$ mass is given by~\eqref{smallfb},
so that non-calculable contributions can indeed be neglected.
In our model, however, the fields $\phi$, $\phi_{1,2}$ must be chosen among
$B$, $\tilde A$, $X$, and $Y_{ij}$, which always results (for $\lambda_I\sim1$)
in  $f_\phi> m_1 m_2$. 
The $\phi_{1,2}$ VEVs are then nonzero, and in order for them
to be small, one must choose $\phi$ to be the highest dimension
field.

\subsection{Supersymmetric minima?}
\label{susyminima}

One can still ask whether, in addition to the supersymmetry-breaking 
minimum near the origin, 
there are any supersymmetric minima or runaway directions elsewhere 
on the moduli space.
As we saw above, supersymmetry can not be restored for finite field VEVs.
However, the  analysis performed so far does not rule out  the 
possibility of runaway 
directions with supersymmetric solutions at infinity, 
because the K\"ahler potential written in terms of the gauge invariants 
can become singular at the boundary of field space. 
In particular, the conclusion that supersymmetry is broken hinges on the assumption that the correct weakly coupled degree of freedom is the baryon, which appears
linearly, as opposed to the antiquark $\barq$.
Thus for large $\barq$ VEVs, this conclusion might not hold.
However, as saw in section~\ref{susybreaking}, the baryon flat direction is lifted.
A nonzero $\barq$ VEV is only possible classically if either $S_M$ or $S_P$ go to infinity.
To study the behavior of the theory at large field VEVs, we will now examine the dynamics along the classical flat directions.
There are three types of singlet flat directions:
\begin{itemize}
 \item $S_M$ gives mass to the vector-like flavors and may play 
a role in the cancellation of the baryon $F$-term.
 \item $S_P$ gives rise to a renormalizable superpotential 
coupling $A\bar Q^2$ in the low-energy theory and may also play a 
role in the cancellation of the baryon $F$-term
 \item $S_{I\ne M,P}$ give rise to NR superpotential 
couplings in the low-energy theory.
\end{itemize}

Let us now examine these directions in the quantum theory, 
starting with $S_M\to\infty$.
Consider the $\barq$ $F$-term equation:
\beq\label{mininfty}
\lambda_M 
 S_M^{ia} Q_i
+ \frac{\lambda_P}{\muv}  S_P^{ab} A\bar{Q}_b 
 +\frac{\lambda_B}{\muv^{2N-3}} \barq^{2N-1}=0\,.
\eeq
The classical potential can be made arbitrarily small if 
the 
flat direction 
$S_M\neq0$
is approached at infinity as\footnote{This requires turning
on additional singlet VEVs for the $F$-term equations to be satisfied.}
\beq
 \label{eqn:clrunaway1}
S_M\rightarrow\infty,~~~
 B=\barq^{2N-1}\neq0,
~~~Q\sim \bar Q^{2N-1}/S_M\rightarrow 0\,.\\
\eeq
While in this ansatz $\bar Q$ is arbitrary, it is constrained to 
be parametrically smaller than $S_M$. 
Thus at energies above 
$\bar Q$ and below $S_M$, 
four of the SU(2$N$) flavors can be integrated out, and the resulting 
effective theory does not have baryons.
This low-energy theory
is the zero-flavor SU(2$N$) with an antisymmetric tensor 
and $2N-4$ anti-fundamentals coupled to some singlets. 
The dynamical superpotential of the  effective theory  
 depends on a power of $S_M$ with a specific exponent. 
In s-confining theories this exponent is always greater than $1$, 
so that the potential is a monotonically increasing function of 
$S_M$. Thus the $S_M$ flat direction is lifted non-perturbatively.

In contrast, the $S_P$ flat direction leads to a runaway in the quantum 
theory. To see this, note that it also can be approached at non-vanishing 
baryon VEV as\footnote{In this case too, other VEVs must be turned on
for the remaining $F$-term equations to be satisfied.}
\begin{equation}
  B=\barq^{2N-1}\neq0,
~~~A\sim \bar Q^{2N-2}/S_P\rightarrow 0\label{clrunaway2}\,.
\end{equation}
However, unlike in the previous case, the $S_P$ VEV only generates 
a renormalizable Yukawa coupling $A\bar Q^2$ of order $\vev{S_P}/\muv$ 
in the effective theory, while the light degrees of freedom of  this 
effective theory remain unchanged. 
As a result, the baryon direction still exists in the field space of 
the low energy theory. 
Studying the theory in the regime
\begin{equation}
 \Lambda\ll \bar Q\ll S_P\ll \muv\,,
\end{equation}
both the gauge 
and the Yukawa interactions are 
weak,
and the potential is calculable in terms of the elementary degrees of freedom. 
In particular, the potential along~(\ref{clrunaway2}) 
can be made arbitrarily small and there is a runaway towards a 
supersymmetric vacuum at infinity (beyond the range of validity of the theory).
Still, 
the supersymmetry-breaking minimum we found earlier is not destabilized 
since near the origin the Coleman-Weinberg potential generates a positive 
mass for $B$,   while $S_P$ is massive due to the non-perturbative dynamics. 
We therefore have a local supersymmetry-breaking minimum at the
origin, and a runaway direction towards zero potential for 
$B\gg\Lambda$, $S_P\to\infty$, with the two regions separated by a 
non-calculable potential barrier.

Finally,  we do not expect any dramatic effect on the dynamics 
along the directions where only  the $S_{I\neq M,P}$ VEVs are turned on.
An $S_{I\neq M,P}$ VEV below $\muv$ simply gives rise to an 
irrelevant superpotential coupling, such as, for $S_A$, $A^N$. 
The analysis of section~\ref{susybreaking} remains valid in the regime 
of validity of the effective theory, $S_I<\muv$.

\subsection{An \headmath{R}-breaking example}
The IR theory described above preserves an $R$ symmetry which remains unbroken 
in the supersymmetry-breaking vacuum. 
This follows from the fact that this theory is an O'Raifeartaigh model
with all $R$ charges being 0 or 2~\cite{Shih:2007av}, and with only a single 
modulus of charge 2~\cite{Evans:2011pz,Shadmi:2011mt}.
One can easily modify the model in order to obtain spontaneous $R$-symmetry
breaking, following~\cite{Shih:2007av}.
As an example, starting from the superpotential~\eqref{w00}, 
we can set $\lambda_A=0$, and add the 
term\footnote{This term breaks the anomaly free 
R-symmetry defined in table~\ref{mattercontent}. 
However, there still exists an anomalous R-symmetry under which the 
tree-level superpotential is invariant. 
As shown in \cite{Goodman:2011jg}, this anomalous R-symmetry 
becomes an accidental R-symmetry of the IR description.}
\beq
\delta W_0= \frac{\lambda_{Asq}}{\muv^{2N-3}}   (A^N)^2\,,
\eeq
which becomes a quadratic term in $\tilde A$ in the IR.
Intergrating out $M$, $P$ and the corresponding singlets as before, 
as well as $Y$ and $S_Y$, one is left with the superpotential
\beq
 W= B (\tilde{A} X +f_B) + m_X S_X X +m_{Asq}  \tilde{A}^2\,,
\eeq
where $m_{Asq} =\lambda_{Asq} \ruv^{2N-3}\Lambda$,
which reproduces the simplest R-breaking model of~\cite{Shih:2007av}.
As discussed in~\cite{Shih:2007av}, this model has a runaway direction
along $S_X$, but for $f_B< m_X m_{Asq}$, one can get a local stable
minimum near $B=0$. This condition requires some tuning of the parameters.
Specifically, we need to take $\lambda_B< \ruv^N$.

\section{The odd-\headmath{n} single anti-symmetric tensor SU(\headmath{n}) theories}
\label{sec:odd}
It is straightforward to repeat the above analysis for the
s-confining SU($2N+1$) theories with a single anti-symmetric tensor.
Again, these theories have 4 ``extra'' flavors. Their matter content and
symmetries are given in Table~\ref{mattercontent2}.
As we will see, the construction of a supersymmetry breaking
IR model is even simpler in this case.
\begin{table}
\begin{tabular}{|>{$\displaystyle}c<{$}|>{$\displaystyle}c<{$}|>{$\displaystyle}c<{$}|>{$\displaystyle}c<{$}|>{$\displaystyle}c<{$}|>{$\displaystyle}c<{$}|>{$\displaystyle}c<{$}|>{$\displaystyle}c<{$}|}
\hline
 &SU(2N+1)&SU(2N+1)&SU(4)&U(1)^\prime&U(1)&U(1)_R&{\rm dim}\\
\hline
\vphantom{\Yasymm^2}
A & \Yasymm &1&1&0&4&0&1\\
\bar Q& \Yafund &\Yfund & 1& 4& 0&2/(2N+1)&1\\
Q&\Yfund &1&\Yfund&-2N-1&-2N+1&0&1\\
\hline
\hline
\vphantom{\Yasymm^2}
M_{ia}\sim(Q_i\bar Q_a)&&\Yfund&\Yfund&3-2N&-2N+1&2/(2N+1)&2\\
P_{ab}\sim(A\bar{Q}^2)&&\Yasymm&1&8&4&4/(2N+1)&3\\
\hline
\vphantom{\Yasymm^2}
Y_i\sim(A^N Q)&&1&\Yfund&-2N-1&2N+1&0&N+1\\
\bar Y^i\sim(A^{N-1} Q^3)&&1&\Yafund&-6N-3&-2N-1&0&N+2\\
B\sim(\bar Q)^{2N+1}&&1&1&8N+4&0&2&2N+1\\
\hline
\hline
\vphantom{\Yasymm^2}
S_M^{ia}&&\bar{\Yfund}&\bar{\Yfund}&-3+2N&2N-1&2-2/(2N+1)&1\\
S_P^{ab}&&\bar{\Yasymm}&1&-8&-4&2-4/(2N+1)&1\\
S_Y^i&&1&\bar{\Yafund}&2N+1&-2N-1&2&1\\
{S_{\bar Y}}_i&&1&\bar{\Yfund}&6N+3&0&2&1\\
\hline
\end{tabular}
\caption{Matter content of the $SU(2N+1)$ model. 
The quantum numbers for 
charged elementary fields are given in the top part, 
for gauge invariant composites in the middle part, 
and for gauge singlets in the bottom part.}
\label{mattercontent2}
\end{table}
Coupling singlet fields to all the gauge invariants apart from $B$ 
and writing down the full superpotential,
\beqa\label{exactwodd}
 W&=& \left[Y M^3 P^{N-1} + \bar Y M P^N +  
m_M  S_M^{ia} M_{ia} + m_P S_P^{ab} P_{ab}
\right]\nn\\
&+&\left[ B \left(\bar Y Y +f_B\right) 
+m_Y S_Y Y +m_{\bar Y} S_{\bar Y} \bar Y\right] 
\,,
\eeqa
we see that once again
the superpotential factorizes into two parts,
with $M$ and $P$ appearing only in the NR terms of the dynamical 
superpotential.
The remaining fields give rise to an O'Raifeartaigh model
in the IR. 
There is a calculable, supersymmetry breaking vacuum 
near the origin.
As in the even-$n$ case, with no tuning of order one couplings
we have $f_B>m_Y m_{\bar{Y}}$, and at the minimum all VEVs are zero
apart from
\beq
Y\sim \ruv^{N-1/2}\Lambda\,,~~~\bar Y\sim \ruv^{N-3/2}\Lambda\,.
\eeq

Just as for the even-$n$ case, the $S_M$ direction is lifted 
quantum mechanically, but there is a runaway direction along $S_P\neq0$.

\section{Other generalizations}
\label{sec:general}
We can generalize our construction to almost all
the s-confining theories. Here we will briefly study a few examples.

Consider $SU(N)$ SQCD with $F=N+1$ flavors~\cite{Seiberg:1994bz}.
The gauge invariants of the theory are the mesons $M_{Ia}=Q_I\barq_a$,
the baryons $B^I={(Q^N)}^I$, and the anti-baryons $\bar B^a={(\barq^N)}^a$,
with $I,a=1,\ldots,F$. We add singlets for all of these gauge invariants
apart from $B^F$. 
We also add the superpotential\footnote{A slightly different deformation of 
s-confining SQCD was considered in~\cite{Goodman:2011jg}. 
There the superpotential was designed to lead to R-symmetry breaking. 
On the other hand, our goal here is to illustrate the result of lifting 
all but one gauge invariant moduli through couplings to singlets.},
\begin{equation}
\label{sqcdW}
W_{{\mathrm class}}=\lambda_S S_{Ia} M_{Ia} +
\frac{\bar\lambda_X }{\muv^{N-2}} \,\bar X_a \bar B^a
+  \frac{\lambda_X}{\muv^{N-2}}\,  X_i B^i
+ \frac{\lambda_B }{\muv^{N-3}}\,B^F\,.
\end{equation}
where $i=1,\ldots,N$, and we wrote the superpotential in terms of invariants
for brevity.
Once again, the only classical flat directions preserved by this superpotential
are parameterized by the gauge singlets
$X$, $\bar X$ and $S$.

The IR superpotential,  
written in terms of canonically normalized fields, is given by
\beq
W=  B^I M_{Ia} \bar B^a  + 
m_M \, S_{Fa} M_{Fa} + \bar m_B\,\bar X_a \bar B^a  
+m_M S_{ia} M_{ia} +  m_B X_i B^i+B^Ff_B \,,
\eeq
with $f_B\sim \lambda_B \ruv^{N-3}$, $m_M\sim\lambda_S \Lambda$, 
$\bar m_B\sim\bar\lambda_X \ruv^{N-2} \Lambda$, and $m_B\sim\lambda_X \ruv^{N-2}$.
Here we omitted  the non-renomalizable $\det M$ term.
Near the origin, the non-perturbative dynamics marries the singlets with the 
gauge-invariant composites.
The mesons $M_{ia}$ and their singlets $S^{ia} $ get mass of order $\Lambda$,
while  $X_i$, $B^i$ get mass $\sim\ruv^{N-2} \Lambda$.     
Since the mesons $M_{ia}$ are stabilized near the origin, the $\det M$ 
term in the dynamical superpotential is irrelevant near the minimum. 
In the IR one then has an \oraf~model 
and the 
Coleman-Wienberg potential stabilizes $B^F$ at the origin.
Note that in this case, we need to take some couplings in~(\ref{sqcdW}) 
to be small in order to get a calculable minimum.
With all couplings of order 1, $f_B > m_M \bar m_B$ and the $\bar B$ VEVs
turn out to be large. 
Choosing $\lambda_B/(\lambda_S \bar \lambda_X)<\ruv$, we have instead
$f_B < m_M \bar m_B$, so that all the fields are stabilized at the origin.

One can check that
in this case, all the flat directions are lifted quantum mechanically.

Another easy to analyze example is  the $SU(5)$ model with three 
``generations'' of 
an anti-symmetric tensor and one anti-fundamental. 
The matter content of the model is presented in Table~\ref{sufivetable}.
\begin{table}
\begin{tabular}{|>{$\displaystyle}c<{$}|>{$\displaystyle}c<{$}|>{$\displaystyle}c<{$}|>{$\displaystyle}c<{$}|>{$\displaystyle}c<{$}|>{$\displaystyle}c<{$}|>{$\displaystyle}c<{$}|} \hline
 &SU(5)&SU(3)&SU(3)&U(1)&U(1)_R&{\rm dim}\\
\hline
\vphantom{\Yasymm^2}
A & \Yasymm &\Yfund&1&1&0&1\\
\bar Q& \Yafund &1&\Yfund & -3& 2/3&1\\
\hline
\hline
\vphantom{\bar Q^{Q^{Q^a}}}
X=A\bar Q^2&1&\Yfund&\Yafund&-5&-4/3&3\\
Y=A^3\bar Q&1&\Yadjoint&\Yfund&0&2/3&4\\
Z=A^5&1&\Ysymm&1&5&0&5\\
\hline
\end{tabular}
\caption{Matter content of the $SU(5)$ model. }
\label{sufivetable}
\end{table}
It was shown in~\cite{Lin:2011vd} that the tree-level superpotential 
given by the sum of all the gauge invariant moduli leads to ISS type 
metastable SUSY breaking. 
We will instead add singlets so that the full superpotential is
\begin{equation}
 W=\frac{1}{\Lambda^{12}}\left(XYZ+Y^3\right)+S_Y Y+S_Z Z + X\,,
\end{equation}
giving an \oraf~model in the IR with a supersymmetry breaking
minimum near the origin.
Again, there are runaway directions along $S_Y\neq0$ and $S_Z\neq0$
for large $A$ VEVs.

One can similarly repeat this construction for theories  3.1.6 or 3.1.9 
of~\cite{sconf}.

\section{Conclusions}
We presented a simple recipe for obtaining effective \oraf~models
in the IR from s-confining theories coupled to singlets.
This construction results in calculable, local supersymmetry-breaking
minima near the origin in many of the s-confining SU($N$) theories.

It is important to note that our construction relies on the presence of 
one or more dynamically-generated terms that are marginal in the infrared. 
Thus not all s-confining theories can lead to SUSY breaking in a non-trivial 
way. For example,  the $SU(7)$ theory with two fields in the symmetric- and 
six fields in the anti-symmetric representation has two moduli, and its 
dynamical superpotential is {\sl quartic} in the moduli~\cite{sconf}. 
Thus it does not give rise to an \oraf~model in the IR.
Indeed the superpotential
\begin{equation}
 W=\frac{1}{\Lambda^{13}}N^2H^2+N+SH\,,
\end{equation}
where, following \cite{sconf}, $H$ and $N$ denote the $SU(7)$ composites 
while $S$ is the gauge singlet,
has a runaway direction along which 
$H\rightarrow 0$ while $S, N\rightarrow\infty$.

The recipe we described can be easily applied to the SO and SP s-confining
theories. 
It would be interesting to generalize it to theories with weakly coupled
IR duals as well.

\section{Acknowlegments}
The research of Y.~Shadmi was supported by
the Israel Science Foundation (ISF) under grant No.~1155/07, 
and by the United States-Israel
Binational Science Foundation (BSF) under grant No.~2006071.
Y. Shirman was supported by the National Science Foundation under 
grant PHY-0970173.


\end{document}